\documentclass[british,english,5p,number,sort&compress, numafflabe, longtitle, times]{melsarticle}
\usepackage{color}
\usepackage{float}
\usepackage{textcomp}
\usepackage{amstext}
\usepackage{graphicx}
\makeatletter

\providecommand{\tabularnewline}{\\}

\usepackage{hyperref}
\usepackage[all]{hypcap}
\usepackage[left]{lineno}
\usepackage{epsfig}
\usepackage{graphicx}

\@ifundefined{showcaptionsetup}{}{%
 \PassOptionsToPackage{caption=false}{subfig}}
\usepackage{subfig}
\makeatother
\usepackage{babel}
\begin{document}

\title{\textcolor{black}Determination of the muon charge sign with \\
the dipolar  spectrometers of the OPERA experiment}

\author[a]{N.~Agafonova} 
\author[b]{A.~Aleksandrov}
\author[c]{A.~Anokhina}
\author[d]{S. Aoki}
\author[e]{A.~Ariga}
\author[e]{T.~Ariga}
\author[f]{D.~Bender}
\author[g]{A.~Bertolin}
\author[h]{C.~Bozza}
\author[i,g]{R.~Brugnera}
% ...

\address[a]{INR Institute for Nuclear Research, Russian Academy of Sciences RUS-117312, Moscow, Russia}
\address[b]{INFN Sezione di Napoli, I-80125 Napoli, Italy}
\address[c]{SINP MSU-Skobeltsyn Institute of Nuclear Physics, Lomonosov Moscow State University, RUS-119992 Moscow, Russia}
\address[d]{Kobe University, J-657-8501 Kobe, Japan}
\address[e]{Albert Einstein Center for Fundamental Physics, Laboratory for High Energy Physics (LHEP), University of Bern, CH-3012 Bern, Switzerland}
\address[f]{METU Middle East Technical University, TR-06531 Ankara, Turkey}
\address[g]{INFN Sezione di Padova, I-35131 Padova, Italy}
\address[h]{Dip. di Fisica dell'Uni. di Salerno and ``Gruppo Collegato'' INFN, I-84084 Fisciano (SA) Italy}
\address[i]{Dipartimento di Fisica dell'Universit\`{a} di Padova, I-35131 Padova, Italy}
% ...

\fntext[note-cern]{Now at CERN, Geneva, Switzerland}
\fntext[note-KR]{Now at CUP, Institute for Basic Science, Daejeon, Korea.}
\fntext[note-samsung]{Now at Samsung Changwon Hospital, SKKU, Changwon, Korea.}
\fntext[note-mich]{Now at Dipartimento di Fisica, Universit\`{a} degli Studi di Trento}
\fntext[note-pol]{Also at National  Research  Nuclear University MEPhI}
\fntext[note-a]{deceased}

\cortext[cor1]{Corresponding authors\\
Universit\"at Hamburg, Luruper Chaussee 149, 22761 Hamburg, Germany}

\begin{abstract}
The OPERA long--baseline neutrino--oscillation experiment  has observed
the direct appearance of $\nu_{\tau}$ in the CNGS $\nu_{\mu}$ beam.
Two large muon magnetic spectrometers are used
to identify muons produced in the $\tau$ leptonic decay and in $\nu_{\mu}$
CC interactions by measuring their charge and momentum. Besides 
the kinematic analysis of the $\tau$ decays, background resulting from the decay of 
charmed particles produced in $\nu_{\mu}$ CC interactions is reduced
by efficiently identifying the muon track.
A new method for the charge sign determination
has been applied, via a weighted angular matching of the straight track--segments
reconstructed in the different parts of the dipole magnets.
Results obtained for Monte Carlo and real data are presented. Comparison with
a method where no matching is used shows a significant reduction
of up to 40\% of the fraction of wrongly determined charges. \end{abstract}
\maketitle

\section{Introduction}

The OPERA experiment has been designed to observe the direct appearance
of $\nu_{\tau}$ in a $\nu_{\mu}$ beam by resolving the track left
by the short--lived $\tau^{-}$ lepton emitted in CC interactions
\citep{Guler}. After analyzing 
about 90\% of the data 
five candidate events have been
observed~\citep{1stTauPaper,2ndTauPaper,Agafonova:2014bcr,Agafonova:2014ptn,5thtaupaper},
corresponding to a significance of $5.1~\sigma$ \citep{5thtaupaper}
thanks to the low level and the control of the background.  The
largest source of background corresponds to charmed particles produced
in $\nu_{\mu}$ CC interactions, which is suppressed by the
identification of the primary (negative) muon.  In case the charmed
particle decays into a muon, $\nu_{\mu}N \rightarrow c\mu^{-}X$ with
$c\rightarrow\mu^{+}Y$, background may be further reduced by
identifying the secondary muon and determining the positive sign of
its charge.  Therefore identifying either primary or secondary muons
is of utmost importance in order to reduce background processes.
Moreover about one
$\nu_{\tau} N \rightarrow \tau^{-} \nu_{\tau} X$
event with
$\tau^{-} \rightarrow\mu^{-} \bar{\nu}_{\mu}\nu_{\tau}$
is expected to
be observed in OPERA, assuming full $\nu_{\mu}-\nu_{\tau}$ mixing and
$\Delta
m_{32}^{2}=2.44\times10^{-3}\,\textrm{eV\text{\texttwosuperior}
}$\citep{Agashe:2014kda}. One such candidate has been actually found
so far.  An efficient and robust estimator of the muon charge sign, in
particular, together with a controlled estimation of its error is
mandatory on an event--by--event basis.

The use of a magnetic spectrometer enables the measurement of the
particle charge and momentum in high energy \linebreak  physics.
In the OPERA experiment~\citep{Acquafredda:2009zz} a cross--sectional area of the order of 70~m$^2$ with a 
rather uniform field was chosen as a good compromise between needs and costs. 
The CERN to Gran Sasso (CNGS) muon--neutrino beam~\cite{cngs-beam} has an average energy of 17~GeV.
A dipolar magnetic field with a uniform magnetization allowing
sufficient bending over the whole magnet cross--section was developed
and implemented in OPERA.  Two large dipoles corresponding to the
bi--modular setup were realized. The two magnetized iron arms of each
dipole, connected by an upper and a lower yoke, are orthogonal to the
neutrino beam~\cite{spect1,spect2}.  Precision Trackers
(PTs)~\citep{Zimmermann:2006xr} consist of six stations of high
precision drift tubes grouped in three pairs, one pair placed in
front, one in between and one behind the two magnetized arms.  Due to
their very low $Z$--density, each PTs station pair reconstructs straight
track segments, enabling the measurement of the deflection angles of
the muon trajectories in the curvature plane\footnote[1]{The OPERA
  coordinate system is a right--handed Cartesian system, with the $Z$
  axis pointing along the horizontal projection of the neutrino beam
  direction, and the $Y$ axis pointing upward. The bending plane is $ZX$,
  perpendicular to the vertical magnetic field lines.}.  A single
magnetized arm together with the four adjacent PT--stations defines
a Charge Measurement Unit (CMU).  Thus the two spectrometers provide
four CMUs.

The Kalman filter procedure~\cite{Kalman} is used in OPERA for the momentum reconstruction,
while the assessment of the charge of the track is based on a independent technique.
In the procedure used so far, the deflection is measured independently for each CMU,
taking into account the energy losses and the Coulomb scattering~\citep{wonsak2007}
when estimating the track momentum. 
The procedure of the charge--sign determination is referred below as the OPERA Standard Method 
(OSM)~\cite{zimmermann2009}, and it is briefly recalled in Appendix A. It relies on the precision
on the angles of the two track segments as parts of the muon trajectory
at the entry and the exit of the magnet arm.
In this paper an up--to--date way to estimate the charge sign is described,
by appropriately exploiting the combined information of the CMUs.
In the following the new procedure is referred to as the Angular Matching Method (AMM).
Results based on Monte Carlo simulations (MC) and real data are
presented together with an estimation of the level of impurity, defined as the fraction of muons for which the charge 
sign is wrongly determined.

It is worthwhile to mention that in
the single candidate observed so far in the muonic decay channel of the $\tau$~\citep{Agafonova:2014bcr},
the low momentum muon stops in the first magnet arm and
its charge cannot be measured by the PT system. Instead, the hits recorded
by the Resistive Plate Chambers (RPC)  that instrument the magnet arm have been
used to assess the negative sign of the muon charge
with a significance of $5.6\, \sigma$. The
improvement in algorithms using the PT data for charge--sign determination thus
does not affect the analysis of this particular
candidate.

\section{The OPERA detector and its muon spectrometers}

The OPERA detector is located at the \foreignlanguage{british}{INFN Laboratori} Na\-zio\-nali
del Gran Sasso (LNGS) in Italy \citep{Acquafredda:2009zz}. It was
exposed between spring 2008 and December 2012 to the CNGS
$\nu_{\mu}$ beam with an average energy of 17 GeV \textcolor{black}{\citep{Bailey1998}},
providing a baseline length of about 730 km to study neutrino oscillation.
The contaminations in CC interactions of $\bar{\nu}_{\mu}$,
$\nu_{e}$ and $\bar{\nu}_{e}$ relative to $\nu_{\mu}$, amount to 2.1\%, 0.9\% and less than 0.1\%, respectively
\citep{Agafonova:2015neo}. The prompt $\nu_{\tau}$ contamination
is negligible. The collected data correspond to about $18\times10^{19}$
protons on target and a total of 19505 neutrino interactions have
been recorded. 

\begin{figure}[H]
\begin{centering}
\includegraphics[width=0.9\columnwidth]{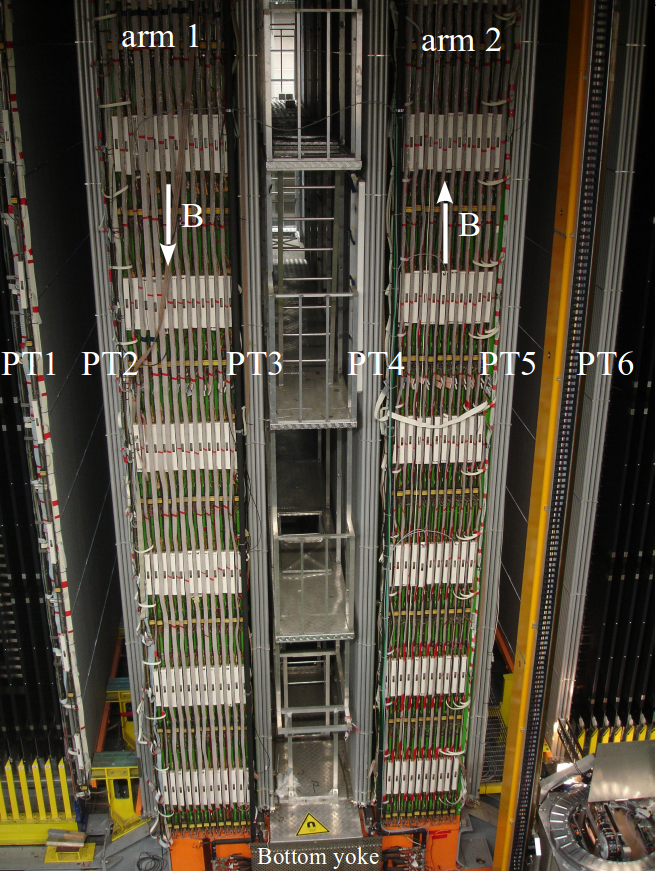}
\par\end{centering}
\caption{Side view of one of the two OPERA spectrometers.}
\label{Flo:OPERA-Photo}
\end{figure}

The OPERA hybrid detector makes use, besides electronic tracking
detectors, of nuclear emulsion to observe with unique space resolution
the production of $\tau$ leptons in $\nu_{\tau}$ CC interactions and
their subsequent decay within a distance of the order of 1 mm.  The
topology of the neutrino interactions is recorded in Emulsion Cloud
Chamber detectors (ECC) with sub-micro\-metric spatial resolution.  The
ECC technique used for the active neutrino target combines nuclear
emulsion films and lead plates to meet the requirement of a
sub-micro\-metric precision over a large volume.
The active neutrino target is made of 150,000 basic units, called
``bricks'' and has a total mass of about 1.2~kt.  The bricks consist
of 56 lead plates of 1 mm thickness interleaved with emulsion
films. They are arranged in 62 vertical walls, each of them followed
by an electronic Target Tracker (TT) plane made of $X$ and $Y$ plastic
scintillator strips having their signal collected by wave--length
shifting fibres and read by multi-anode photomultiplier
tubes~\cite{TT}.  The TT planes are aimed at triggering the data
acquisition and
measuring the trajectories of the charged particles through the
target, selecting the bricks to be extracted from the walls where
neutrino interactions occurred and localizing the area where the
scanning of the emulsion films~\cite{microscopes} has to start in the
search for tracks pointing to the neutrino vertex.
The neutrino target is subdivided into two identical Super--Modules (SM). 

A large magnetized volume is placed downstream of each SM target.
Each spectrometer is an iron dipole, with a uniform magnetic field
over a cross--sectional area of 8 $\times$ 8.75 m$^2$. As shown in
Figure~\ref{Flo:OPERA-Photo}, the magnet consists of two vertical arms
with upper and lower horizontal yokes to close the magnetic field
circuit. The 1.53~T magnetic field has opposite directions in the two
arms. Non--uniformities have been measured to be less than
3\%~\citep{spect2}.  Each magnet arm is segmented in twelve 5~cm thick
iron--slabs, internally instrumented by eleven planes of RPC equipped
with $X$--$Y$ strips for a coarse tracking (spatial resolution
\textasciitilde{}1cm).  Externally to the arms each spectrometer is
equipped by six stations of high precision vertical drift tubes
planes, the PTs~\citep{Zimmermann:2006xr}, for precise muon
tracking. Each PT station consists of 4 layers of drift tubes.  The
non--bending coordinate ($Y$) is coarsely measured by two walls of 
RPC planes, placed upstream of the magnet with the readout strips
inclined by $\pm$~42.6$^{\circ}$ with respect to the horizontal
direction.  They are placed 1~cm upstream and downstream of the first
and second drift tube station, respectively.

\begin{figure*}
\begin{centering}
\includegraphics[width=0.85\textwidth]{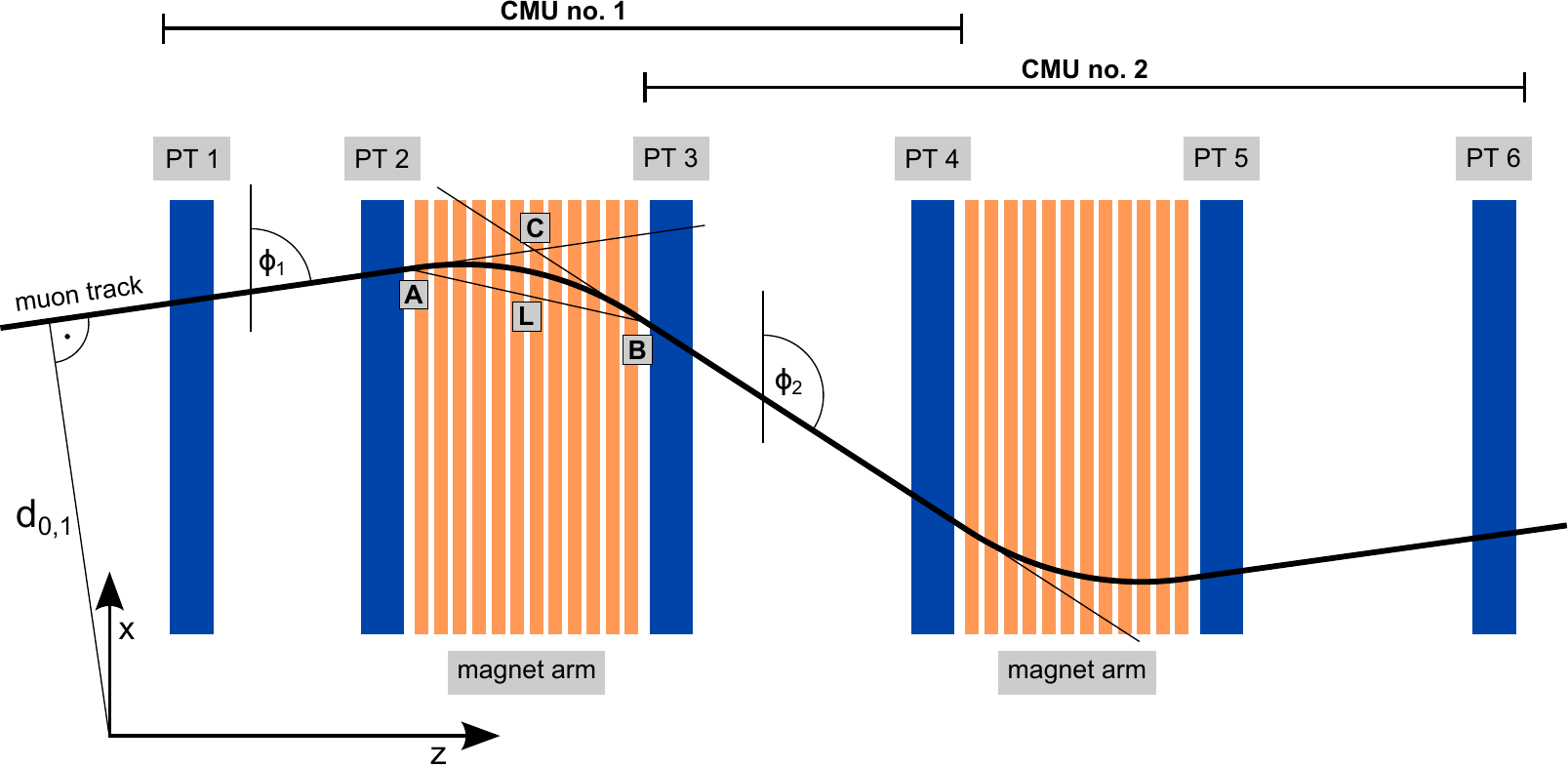}
\par\end{centering}
\caption{Schematic top view of one of the two OPERA spectrometers. The
  six Precision Tracker (PT) stations are used for the track
  reconstruction.  Each Charge Measurement Unit (CMU) delivers a
  measurement of charge/momentum, provided the track is reconstructed
  on both sides of the magnet arm.  PT1--PT4 form the first CMU and
  PT3--PT6 form the second CMU. The charge sign determination in the
  first CMU uses the deflection angle
  $\Delta\phi=\phi_{2}-\phi_{1}$\textcolor{black}{.  The OPERA
    detector has a total of four CMUs, two per spectrometer.}}
\label{Flo:ChargeSignDet}
\end{figure*}

Each spectrometer consists of two CMUs given by the two pairs of PT
stations, upstream and downstream of each magnet arm
(Figure~\ref{Flo:ChargeSignDet}).
The CMUs provide a measurement of the charge and the horizontal
projection of the muon momentum, provided the track is reconstructed
on both sides of the magnet arm. The spatial resolution of the PT is
better than $300\,\text{\textmu m}$ in the horizontal plane
\citep{Zimmermann:2006xr, agafonova2010}.

\section{Monte Carlo simulations\label{sec:Monte-Carlo}}

Two different Monte Carlo simulations based on
\textsc{Geant3}\cite{geant}
have been used. For both types of simulation muon trajectories have
been reconstructed and their momenta evaluated using the full OPERA
analysis chain \citep{StudyInteractions}.

The first MC simulation (MC--I) was used to demonstrate the performances
of AMM as function of the muon momentum. The sample consisted of positive and negative muons emitted
at the \foreignlanguage{british}{centre} of the targets of either SM1 or
SM2, at an angle orthogonal to the drift tubes planes. The momentum
was varied between 1 and 100 GeV/c in steps of 1 GeV/c. The angle
is close to those of most muons produced in CNGS CC neutrino interactions
in the detector. 

The full OPERA simulation chain of the response of the electronic detectors to CNGS neutrinos interactions (MC--II) 
has been used in a second step in view of evaluating the potential gain that the AMM algorithm may provide in the 
muon charge determination. It uses the event generator NEGN~\citep{eventGen} 
developed in the framework of the NOMAD experiment~\citep{NOMAD}.

\section{Methodology}

\subsection{Identification of track inconsistencies\label{sub:Identification-of-Track}}
\label{subsec:id}

\begin{figure}[htbp]
\begin{centering}
\textcolor{black}{\includegraphics[width=3.5cm]{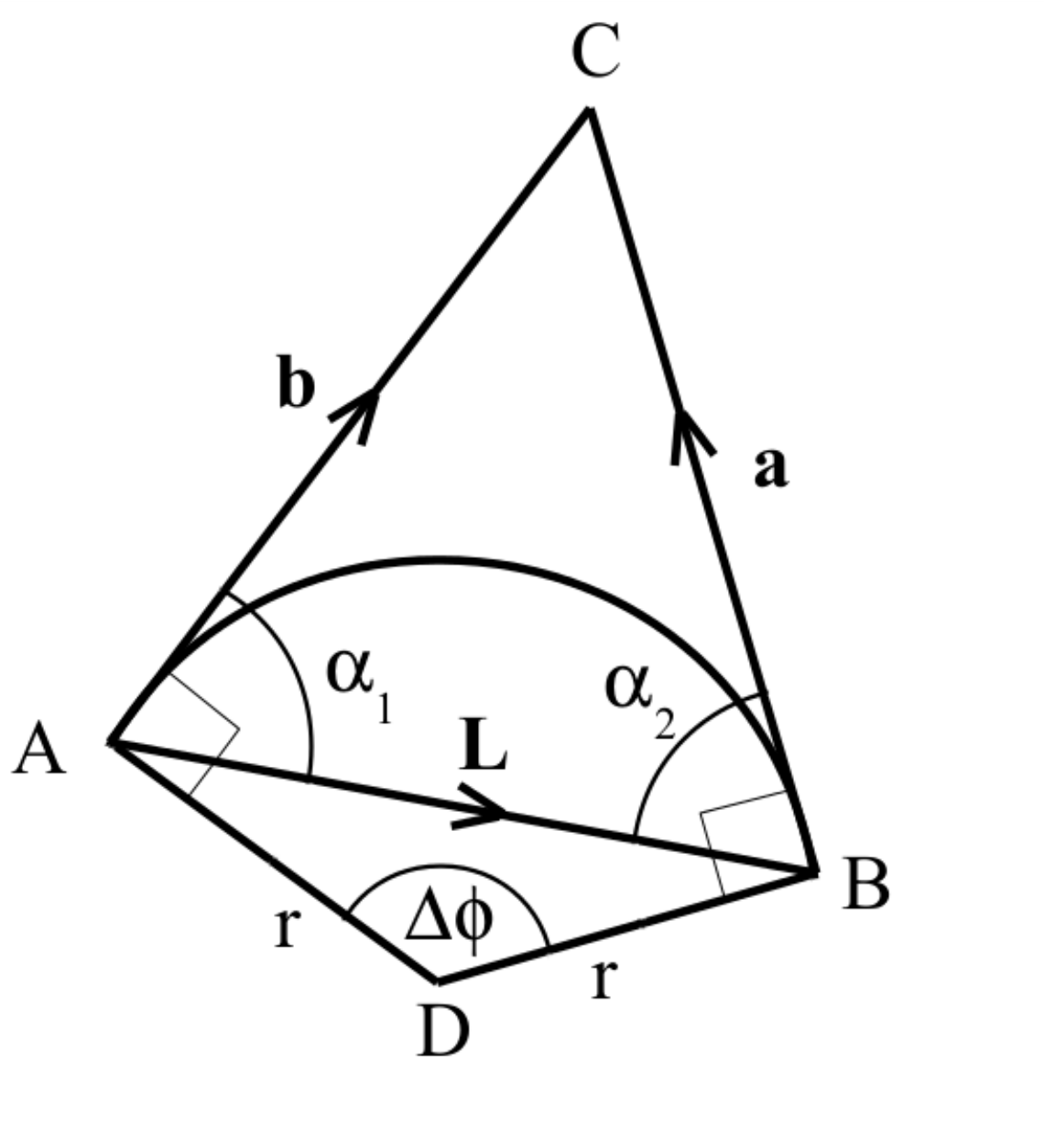}}
\par\end{centering}
\caption{The particle trajectory in the magnetic field is shown (arc of a circle
AB of radius $r$). Vectors $\mathbf{a}$ and $\mathbf{b}$ (intersecting
at C) are tangents to the circular path at the entry (A) and exit
(B) points. A, B and C refer to points in Figure \ref{Flo:ChargeSignDet}.
$\Delta\phi=\phi_{2}-\phi_{1}$ is the deflection angle.}
\label{Flo:triangle}
\end{figure}

  The straight track segments reconstructed by the PT system provide
  track parameters projected in the horizontal plane ($X$, $Z$),
orthogonal to the magnetic field lines that are directed along the $Y$
axis.  For detailed information on track reconstruction in the PT see
\citep{wonsak2007}. Ideally, two segmen\textcolor{black}{ts on each
  side $j=1,2$ of a magnet arm are described by the angles $\phi_{j}$
  that they }enclose\textcolor{black}{{} with axis $X$ and by their
  distance $d_{0,j}$ to the reference frame origin (see Figure
  \ref{Flo:ChargeSignDet}; for reasons of simplicity $d_{0,2}$ is not
  shown in the figure). $\Delta\phi=\phi_{2}-\phi_{1}$ is the
  deflection angle. If energy loss and multiple scattering are
  neglected, a charged particle trajectory in the magnetic field is an
  arc of a circle tangent to both track segments at their }magnet arms
entry/exit points (see Figure \ref{Flo:triangle}). Calling
$\alpha_{j}$ the complement of the angle between the chord joining
these two points and track segment $j$, it follows that
$\alpha_{1}=\alpha_{2}$ and $\Delta\phi=\alpha_{1}+\alpha_{2}$ (see
Appendix B for more details). The relative angular deviation in
CMU$_i$

\textcolor{black}{
\begin{equation}
\Delta\left(\alpha\right)_{rel,i}=\left.2\frac{\alpha_{1}-\alpha_{2}}{\alpha_{1}+\alpha_{2}}\right|_{i}=\left.2\frac{\alpha_{1}-\alpha_{2}}{\Delta\phi}\right|_{i}\label{eq:mark}
\end{equation}
provides a measurement of the mismatch between the two reconstructed
track segments.  Large values of the mismatch between $\alpha_{1}$ and
$\alpha_{2}$ indicate that one or both track segments are
reconstructed with poor precision, in which case they may also be
wrongly associated, yielding to a wrong muon sign determination.
}

In the following a weight is constructed such that
large angular mismatches result in a small weight. 

\begin{figure*}[htbp]
\begin{centering}
\textcolor{black}{}\subfloat[Charge weight principle]{\begin{centering}
\textcolor{black}{\includegraphics[width=1\columnwidth]{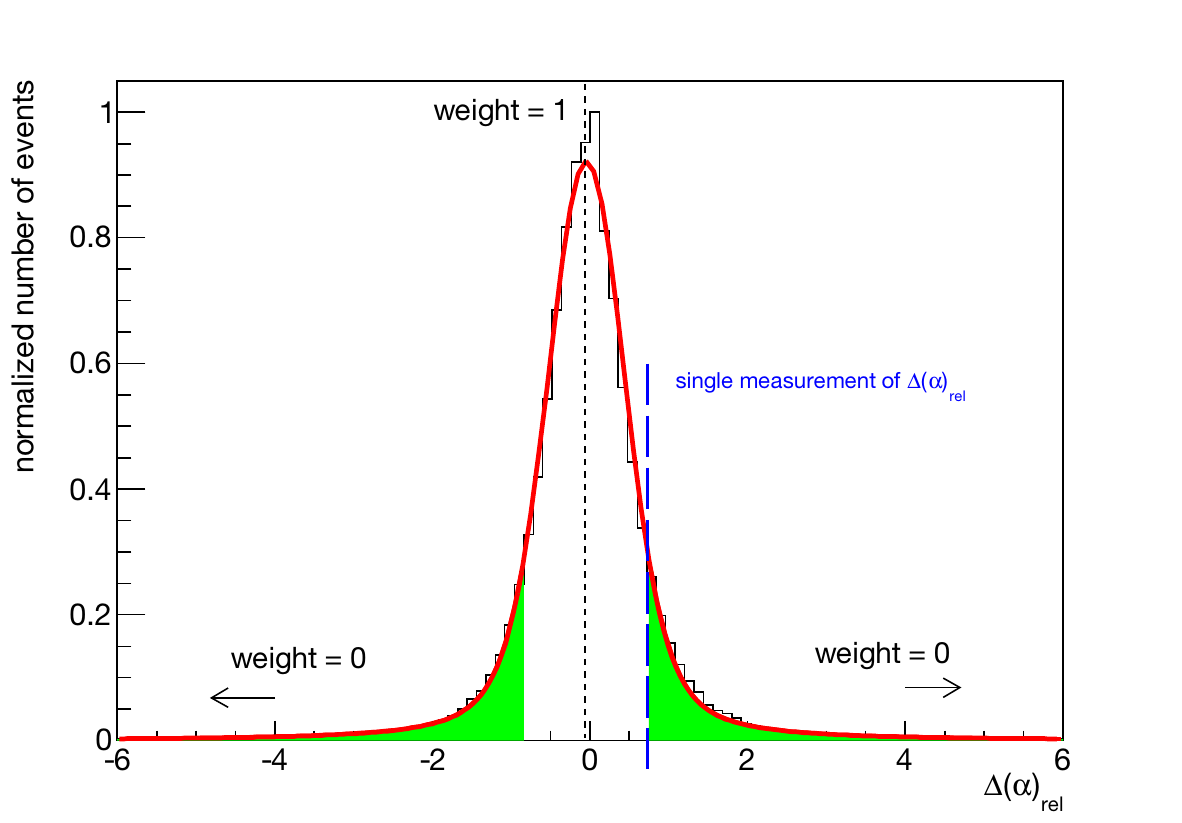}}
\par\end{centering}
\textcolor{black}{\label{wp}}}\textcolor{black}{}\subfloat[Charge weight dependence on momentum]{\begin{centering}
    \textcolor{black}{\includegraphics[width=1\columnwidth]{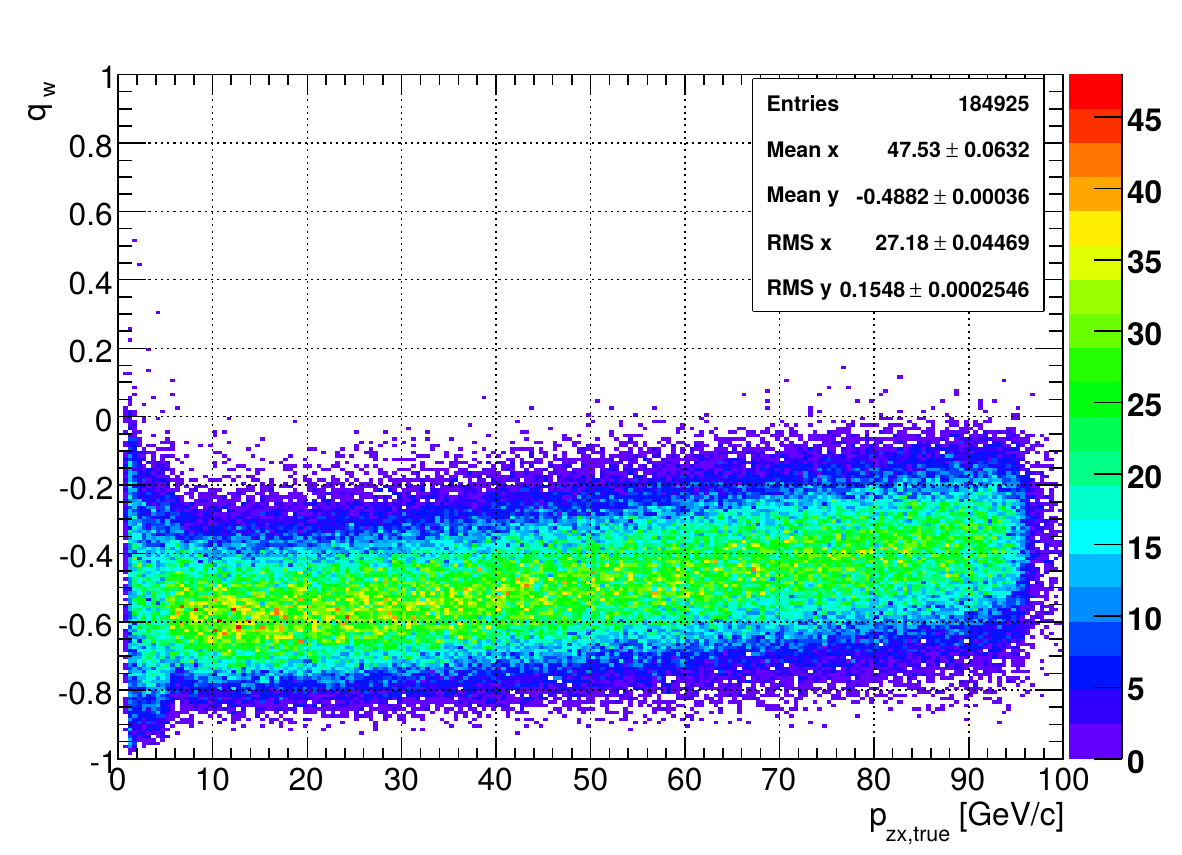}}
\end{centering}
\textcolor{black}{\label{Flo:ChargeWeight_Momentum}}}
\par\end{centering}
\textcolor{black}{\caption{(a): The black histogram shows the distribution of the relative angular
deviations $\Delta\left(\alpha\right)_{rel}$ (Eq. \ref{eq:mark})
and the red curve the Voigtian distribution that best fits to it.
The central value
$\mu$ is indicated by the dashed line. The weight $w_{i}$ attributed
to the charge sign measurement obtained in a single CMU$_i$ is given
by the integral defined in Eq. (\ref{eq:weight_HPT}), (\ref{eq:inLimits1})
and (\ref{eq:inLimits2}). It corresponds to the green area. The
smaller $\Delta\left(\alpha\right)_{rel}$ is, the larger is the weight.
Figure (b) shows, for simulated $\mu^{-}$, the dependence of the
 charge weight estimator $q_{w}$ (Eq. (\ref{eq:w_tot})) on the true Monte
Carlo momentum $p_{{\rm ZX},true}$ projected in the horizontal plane. The
sign of $q_{w}$ fixes the sign of the reconstructed muon charge
(see Section \ref{sub:Charge-Sign-Determination}). For both figures
the simulation method MC--I is used with negative muons (see Section
\ref{sec:Monte-Carlo}), with the difference that for Figure (b) the momentum
is uniformly distributed between 1 and 100 GeV/c instead of in steps
of 1 GeV/c. }}
\textcolor{black}{\label{Flo:WeightPrinciple}}
\end{figure*}

The distribution of the reconstructed $\Delta\left(\alpha\right)_{rel}$
for 200,000 negative muons simulated following MC--I (see Section \ref{sec:Monte-Carlo})
is shown in Figure \ref{wp}. It has been fitted with a Voigtian distribution
$V\left(x\right)$, the convolution of a Gaussian $G\left(x\right)$
and a Lorentzian $L\left(x\right)$ distribution,
\begin{eqnarray}
V\left(x\right) & = & G\left(x\right)\otimes L\left(x\right)\\
 & =A & \frac{1}{\sqrt{2\pi\sigma}}e^{-\frac{1}{2}\left(\frac{x-\mu}{\sigma}\right)^{2}}\otimes\frac{1}{2\pi}\frac{\gamma}{\left(x-\mu\right)^{2}+\frac{1}{4}\gamma^{2}},
\end{eqnarray}
where $\mu$, $\gamma$
and $\sigma$ are obtained from the fit.
\textcolor{blue}{{} }
$A$ is a normalization factor. The central value $\mu$ is
fully compatible with 0. \textcolor{black}{The relative error on the
angle difference given by the precision on its measurement
is well reproduced by a Gaussian distribution except for the long tails
that are described by the convolution with a Lorentzian distribution.}
The quality of the track reconstruction in CMU$_i$ is given by its
weight

\begin{eqnarray}
w_{i} & = & 1-\int_{a_{i}}^{d_{i}}V\left(\Delta\left(\alpha\right)_{rel,i}\right)\textrm{d}\Delta\left(\alpha\right)_{rel,i}.\label{eq:weight_HPT}
\end{eqnarray}
The integration limits are 

\begin{eqnarray}
d_{i} & = & \mu+\left|\Delta\left(\alpha\right)_{rel,i}-\mu\right|\label{eq:inLimits1}\\
a_{i} & = & \mu-\left|\Delta\left(\alpha\right)_{rel,i}-\mu\right|.\label{eq:inLimits2}
\end{eqnarray}
For small values of $\Delta\left(\alpha\right)_{rel,i}$ a weight of \textasciitilde{}1
is attributed to the measurement, while it approaches 0 for large
values of $\left|\Delta\left(\alpha\right)_{rel,i}\right|$ (Figure~\ref{wp}).

\subsection{Charge sign determination\label{sub:Charge-Sign-Determination}}

The OPERA detector consists of four CMU \textcolor{black}{each allowing}
the determination of the deflection angle $\Delta\phi$. For muons
originated in CNGS neutrino interactions, taking into account the polarity
$P_{i}$ of the magnetic field in CMU$_i$, the charge sign $c_{i}$
is given by

\begin{equation}
c_{i}=\frac{\phi_{i2}-\phi_{i1}}{|\phi_{i2}-\phi_{i1}|}P_{i},
\end{equation}
where $\phi_{ij}$ represents the reconstructed angle in front of
($j=1)$ and behind ($j=2$) the magnet arm of CMU$_i$. A charge sign estimator is defined as

\begin{equation}
q_{w}=\frac{1}{n}\cdot\sum_{i=1}^{n}c_{i}\cdot w_{i},\label{eq:w_tot}
\end{equation}
where $n$ denotes the number of used CMUs.\textcolor{black}{{} The sign
  of $q_{w}$ corresponds to the reconstructed particle charge sign.}
Its modulus measures the quality of this determination.
Figure \ref{Flo:ChargeWeight_Momentum}
shows, for simulated $\mu^{-}$, the charge sign estimator $q_{w}$ as a function of the true
MC momentum projected in the horizontal plane $p_{{\rm ZX},true}$\textcolor{black}{. 
Negative muons with $q_{w}>0$ have
their charge misidentified.
The momentum dependence
of the wrong sign determination is small and further discussed in
Section \ref{sec:Charge-Misidentification-Results}. In the range of momentum 
relevant to the physics of OPERA, this can be an important advantage
in comparison with conventional methods in particle physics as these
are strongly momentum dependent (e.g. Kalman tracking \cite{Kalman}).
The statistically optimal behavior of the Kalman fit, or equivalently
a $\chi^2$ fit, is actually valid only within linear approximations
which may not hold in presence of
energy loss and multiple scattering.  Our result shows that a
non--conventional approach, applied to specific cases and to extract
only partial information (charge sign), may be more appropriate.  The
width of the $q_{w}$ distribution is dominated by the measurement
errors on the angles and is essentially momentum independent except at
low momentum where the muon trajectory inside the magnets is modified
by multiple Coulomb scattering.}

\section{Charge \foreignlanguage{british}{misidentification} results\label{sec:Charge-Misidentification-Results}}

\subsection{Performances of AMM\label{sub:Monte-Carlo}}

\begin{figure*}[htbp]
\begin{centering}
\subfloat[]{\begin{centering}
\includegraphics[width=0.5\textwidth]{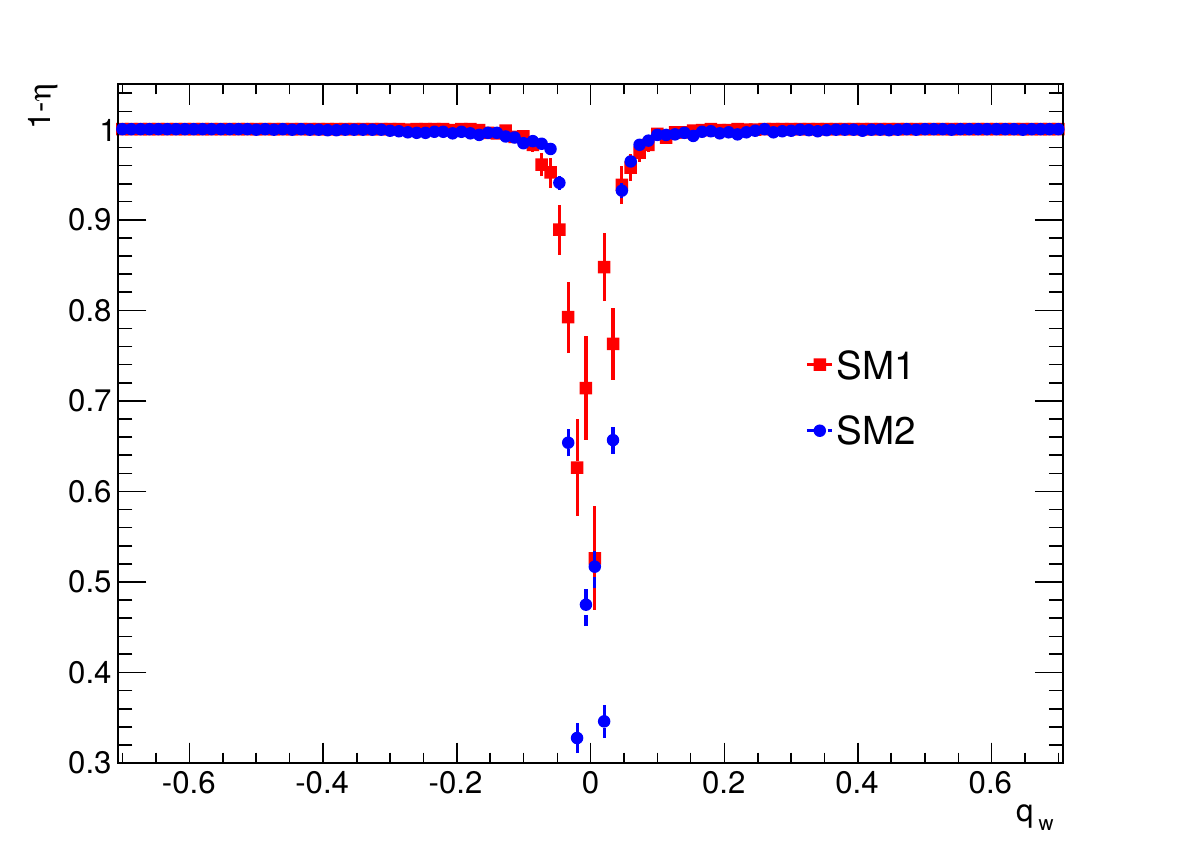}
\par\end{centering}
\label{Flo:Imp_Eff_W_Cut}}\subfloat[]{\begin{centering}
\includegraphics[width=0.5\textwidth]{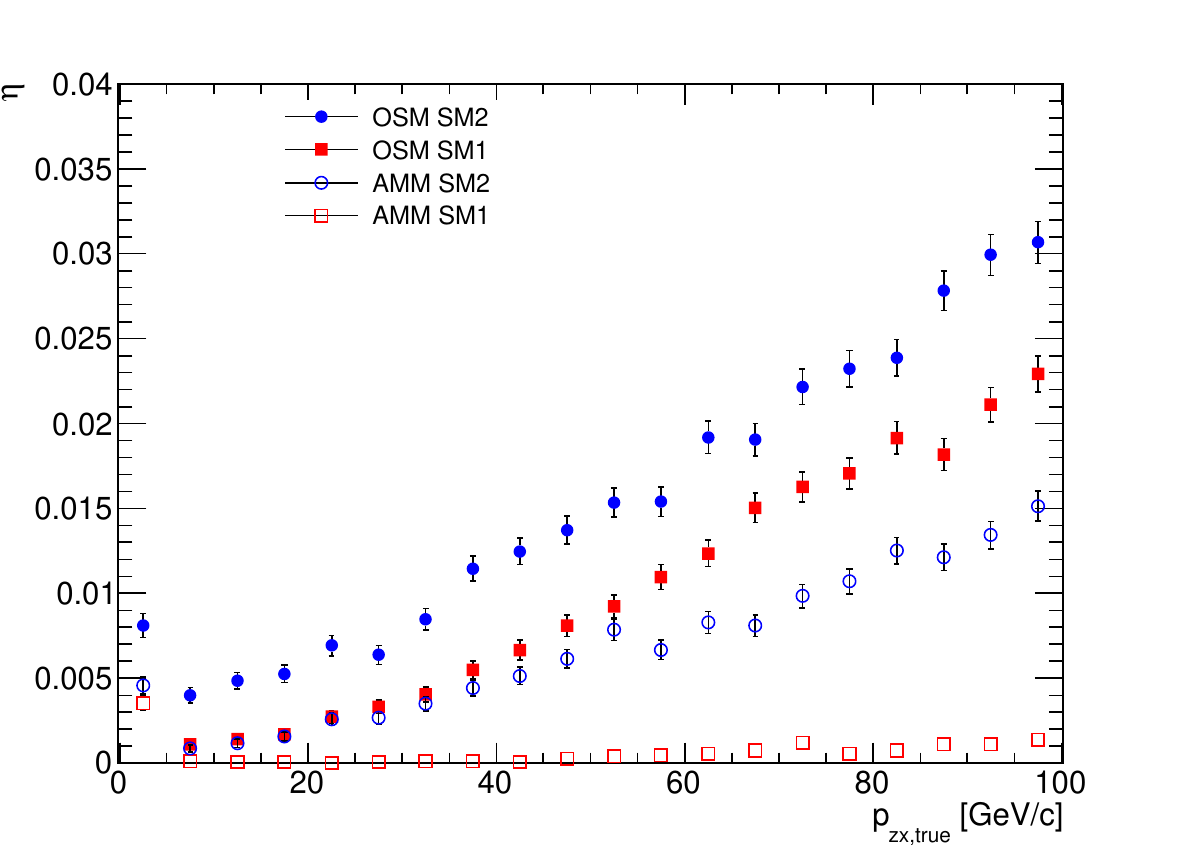}
\par\end{centering}
\label{Flo:impurity_all}}
\par\end{centering}
\caption{Purity $1-\eta$ dependence on the estimator $q_{w}$. The vertex
of the incoming muons was set in the middle of the target section,
in the first (SM1) and in the second supermodule (SM2), respectively.
The impurity $\eta$ dependence on the momentum $p_{{\rm ZX},true}$ is
shown in (b). The simulation method is MC--I.}
\end{figure*}

To demonstrate the potential of the new charge--sign algorithm the
simulation MC--I was used, and the impurity $\eta$ as well as the efficiency
$\varepsilon$ for $\mu^{+}$ and $\mu^{-}$ were calculated. The
impurity is defined by
\begin{equation}
\eta=\frac{n_{w}}{n_{c}}
\end{equation}
where $n_{w}$ is the number of wrong charge sign determinations,
and $n_{c}$ is the total number of charge sign assignments. The efficiency is defined
by $\varepsilon=n/n_{p},$ where $n=n_{c}-n_{w}$ is the number of
correct charge sign assignments and $n_{p}$ the number of possible
charge sign assignments, where at least one CMU is crossed by the
particle. If a cut on the estimator $q_{w}$ is applied the discarded events
are not included in $n_{c}$ and $n_{w}$.  If no cut is applied to $q_{w}$ then $\eta=1-\varepsilon$.

At very low momentum, purity and efficiency
suffer from the multiple Coulomb scattering inside the magnet,
and, at very high momentum, from the finite resolution of the PT.
A cut on the weight $q_{w}$ increases the purity as shown in
Figure \ref{Flo:Imp_Eff_W_Cut} at the cost of a reduction of the
efficiency. The results obtained for the impurity are shown in Table
\ref{compImpAMMOSM} and in Figure \ref{Flo:impurity_all} for both
OSM and AMM. The following observations are made for AMM:
\begin{itemize}
\item Even in the absence of cuts on the weight $q_{w}$, the impurity
remains smaller than 0.1\% at momenta larger than 5 \foreignlanguage{british}{GeV}/c,
if the vertex is set in the target of SM1. In the momentum range considered
in the search for $\nu_{\tau}$ candidate events, $p_{\mu}$< 15~GeV/c,
it does not exceed 0.01\% except for momenta smaller than 5~GeV/c
where it increases to 0.4\% (Figure \ref{Flo:impurity_all}).
\item If the vertex is set in the target of SM2 the impurity
  increases, but still does not exceed 1.6\% at high momentum and
  0.2\% for 5~<$ p_{\mu} $<15~GeV/c. For $p_{\mu}$ < 5 GeV/c it
  increases to 0.5\% (Figure \ref{Flo:impurity_all}).
\end{itemize}
A cut on the weight $\left|q_{w}\right|>0.1\,\left({\rm or}\; 0.2\right)$
removes most of the impurities if the vertex is set in the target
of SM1 even at very low or very high momentum at the cost of a reduction
in the efficiency as large as 10 \% at very high momentum. For $\textrm{5\,\ GeV/c}<p_{\mu}<\textrm{15\,\ GeV/c}$,
however, the efficiency remains larger than 99.7\% but it falls to
99\% (or 96\%) for $p_{\mu}$ < 5 GeV/c when the cut on the weight is applied. 

\begin{table}[h]
\caption{Comparison between the total impurities, averaged over the momentum
range 1--100 GeV/c, obtained with  new AMM and  OSM. }

\begin{centering}
\begin{tabular}{ccc}
Vertex position & $\eta_{OSM}\left(\textrm{all}\right)\left[\%\right]$ & $\eta_{AMM}\left(\textrm{all}\right)\left[\%\right]$\tabularnewline
\hline 
target of SM1 & $1.35 \pm 0.02$ & $0.06 \pm 0.01$\tabularnewline
target of SM2 & $1.55 \pm 0.02$ & $0.69 \pm 0.01$\tabularnewline
all & $1.45 \pm 0.01$ & $0.37 \pm 0.01$\tabularnewline
\hline 
\end{tabular}
\par\end{centering}

\label{compImpAMMOSM}
\end{table}
 The already small fraction of wrong charge sign determination obtained
with  OSM is reduced by an order of magnitude when the vertex is
placed in the target of SM1 and by a factor of 4 on average. The potential
impact of this improvement on the physics results of OPERA is discussed
in Section \ref{sec:Conclusion}.

\subsection{Simulation of CNGS neutrinos CC interactions}

\begin{figure}[t]
\begin{centering}
\includegraphics[width=1\columnwidth]{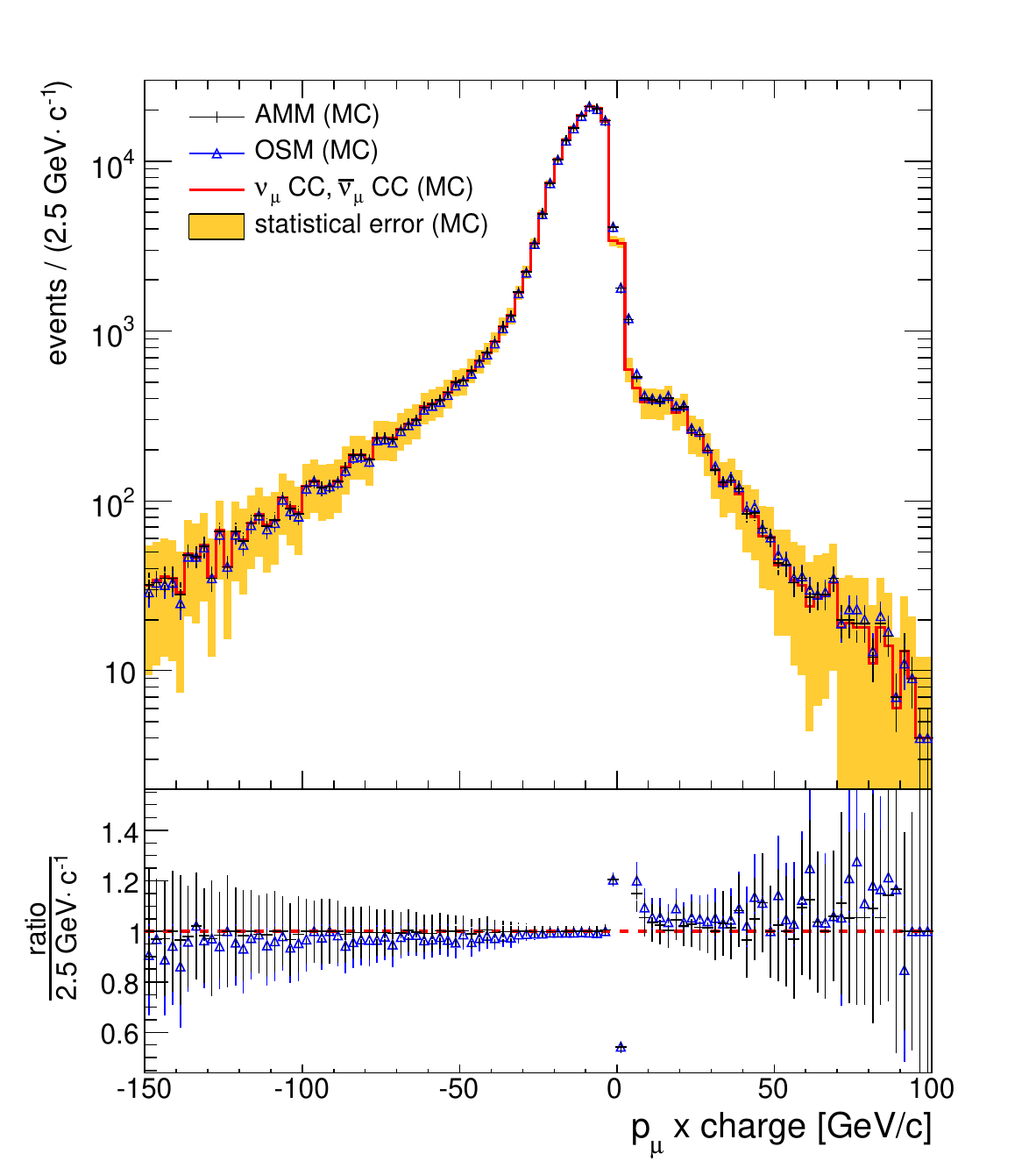}
\par\end{centering}

\caption{Top: reconstructed (anti--)muon momentum multiplied with the sign of
its charge for simulated CNGS CC neutrino interactions (MC--II). The
charge is obtained by the truth--MC (red histogram), and the charge reconstructed
with either OSM (blue histogram with triangle) or AMM (black histogram
with vertical mark). For visibility, the statistical error (orange
band) on the first spectrum (truth--MC sign) has been multiplied by
3. Bottom: ratios between the truth--MC charge sign and the reconstructed
charge signs obtained by either OSM or AMM. }

\label{Flo:MomentumSpectrum}
\end{figure}

A sample of CNGS beam CC neutrino interactions has been generated in
the detector target, using the full OPERA simulation chain MC--II
described above \cite{StudyInteractions}. Figure
\ref{Flo:MomentumSpectrum} shows the distribution of the product of
the reconstructed muon momentum $p_{\mu}$ and its charge in a range
extending from $-$150 GeV/c to 100 GeV/c that includes essentially all
events. The fraction of $\mu^{+}$ in the total sample of muons
estimated with AMM and OSM are (3.7~\textpm{}~0.1(stat.))\% and
(3.9~\textpm{}~0.1(stat.))\%, respectively. The estimated fraction of
$\mu^{+}$, mainly emitted in CC interactions of $\bar{\nu}_{\mu}$ from
beam contamination, can only be biased towards values larger than the
MC--truth, (3.4~\textpm{}~0.1(stat.))\%.  The fraction of incorrect
charge assignments is reduced by \textasciitilde{}40\%, from 0.5\% to
0.3\%, with AMM.

Figure 7 shows the muon charge determination impurity as a function
of the momentum in the range 2~<~$p_{\mu}$~<~150~GeV/c obtained
with both methods. If a maximum momentum cut is applied at 15 GeV/c,
the impurities obtained with OSM and AMM are respectively (1.06~\textpm{}~0.04~(stat.))\%
and (0.62~\textpm{}~0.03 (stat.))\%, i.e. a reduction of $\sim$~40\%.
If a cut $\left|q_{w}\right|$>~0.1 is applied to the charge determination, the impurity is reduced
by a factor of 3, while the efficiency is reduced to \textasciitilde{}86\%.
\begin{figure}
\begin{centering}
\includegraphics[width=1\columnwidth]{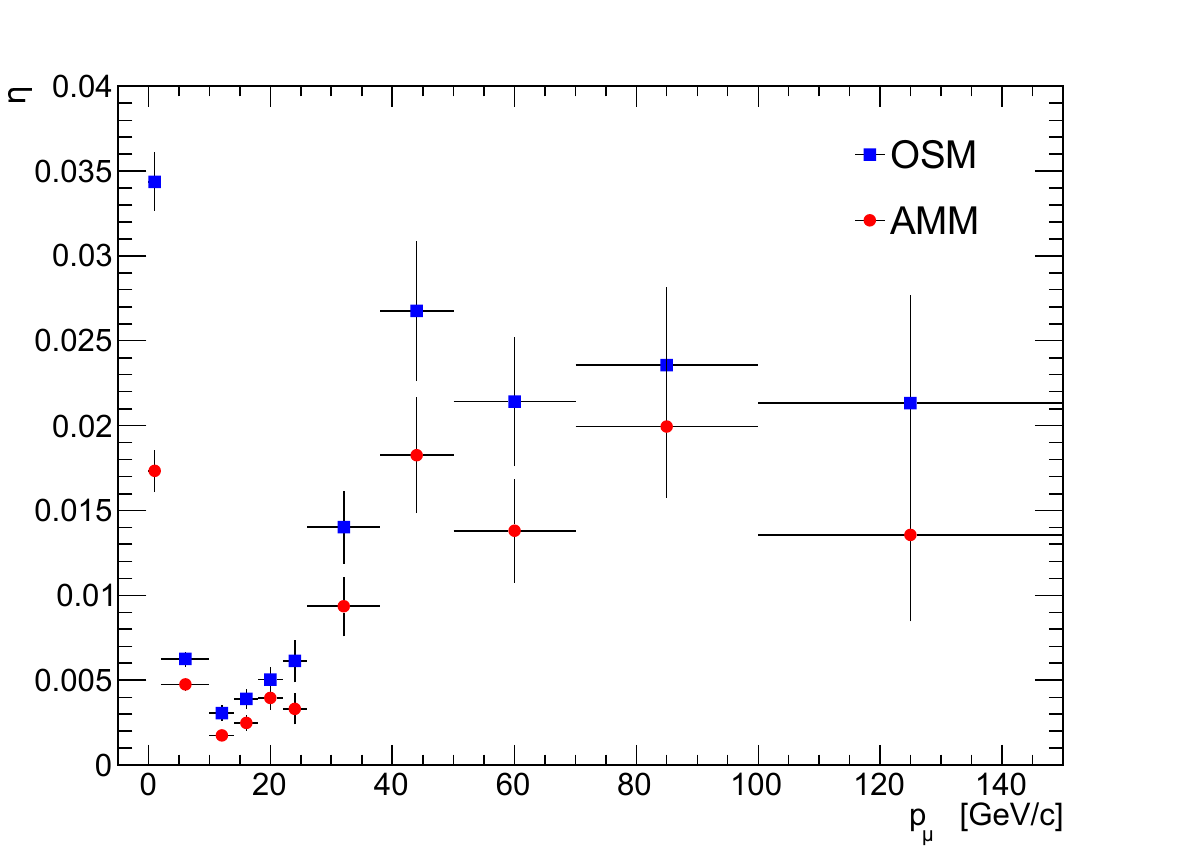}
\par\end{centering}

\caption{Dependence on the muon momentum $p_{\mu}$ of the impurity $\eta$
on the charge sign determination for simulated CNGS CC neutrino interactions
(MC--II).}

\label{Flo:Impurity_realisticMC}
\end{figure}

\section{Charge misidentification for real data}

In order to test the new method with real data, two investigations
were carried out.

\begin{figure}
\begin{centering}
  \includegraphics[width=1\columnwidth]{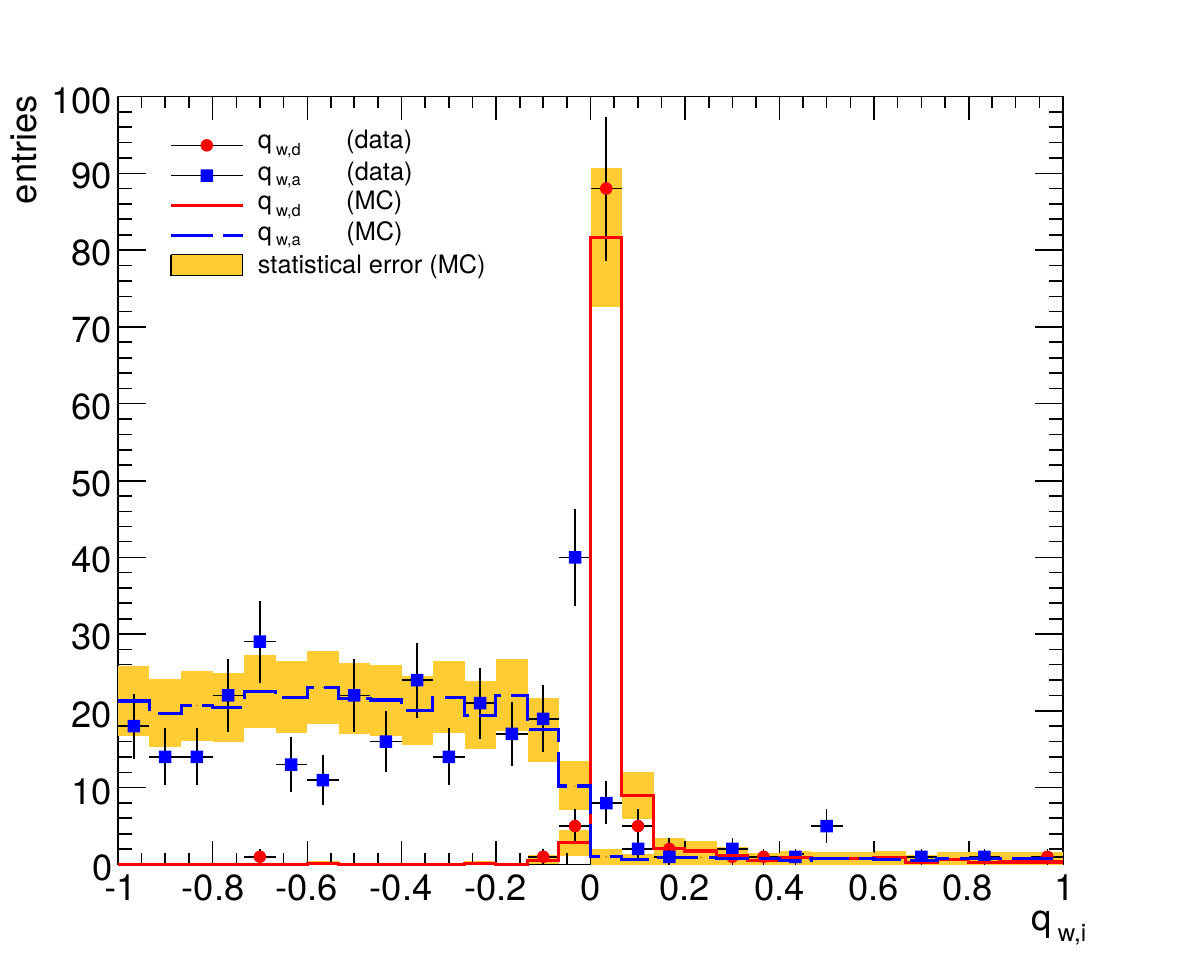}
\par\end{centering}

\caption{Real data and MC comparison: Tracks that propagate through
the all four charge measurement units and where three stations are in
agreement and one station is in disagreement are used. $q_{w,a}$ is the
weight of the stations in agreement while $q_{w,d}$ is the weight
of the station in disagreement. This station is expected to have more
often delivered a wrong charge sign, and indeed it yields a small
absolute value of the weight.\protect \\
}

\label{Flo:Test with real data_agree}
\end{figure}

Neutrino interactions were selected in which the muon propagates through
both spectrometers (number of used CMU \linebreak  $n_{CMU}=4$)
and has one charge measurement in disagreement with the other three
measurements. The single sign measurement is more likely to be incorrect
than the other three measurements. Figure \ref{Flo:Test with real data_agree}
shows as an example the weight of the three stations in agreement
$q_{w,a}$ and the weight of the one station in disagreement, $q_{w,d}$.
Compared to $\left|q_{w,a}\right|$, $\left|q_{w,d}\right|$
is generally very small, $|q_{w,d}|<0.2$. 

\begin{figure}[!t]
\begin{centering}
\subfloat[No cut.]{\begin{centering}
\includegraphics[width=1\columnwidth]{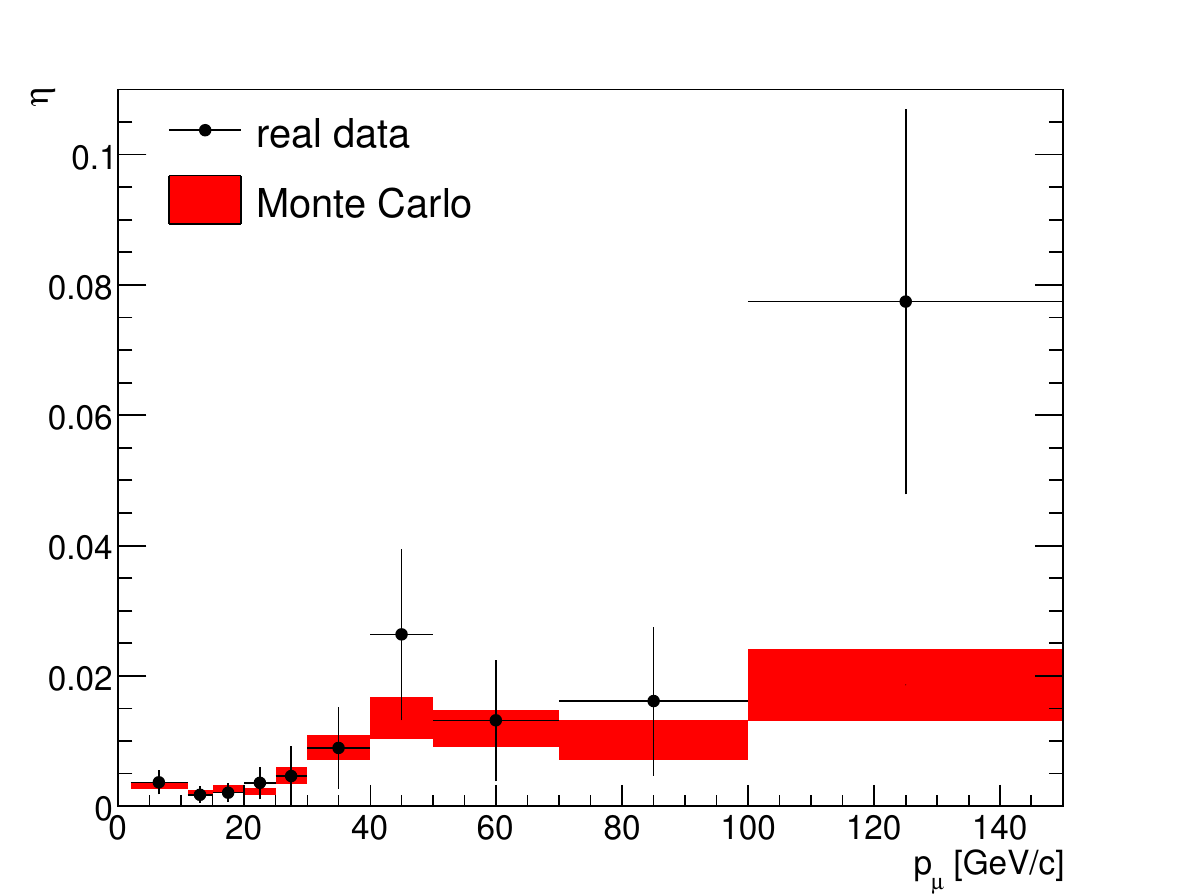}
\par\end{centering}

\label{Flo:NoWeightCut}}\\
\subfloat[Cut at $\left|q_{w}\right|>0.1$]{\begin{centering}
\includegraphics[width=1\columnwidth]{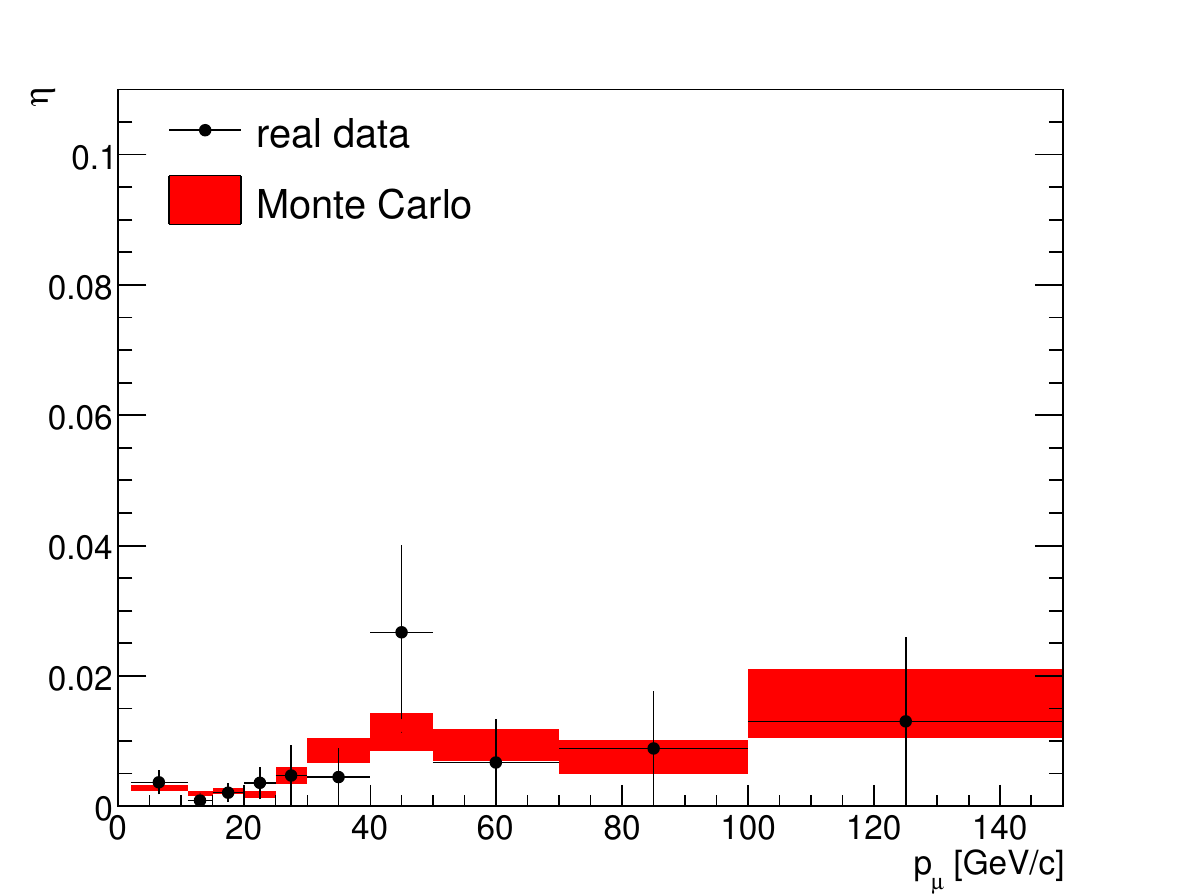}
\par\end{centering}

\label{Flo:WeightCut}}
\par\end{centering}

\caption{Evaluation of the impurity $\eta$ with real data compared to the
MC expectation as a function of the muon momentum $p_{\mu}$ for one
SM using AMM (Eq. \ref{eq:w_tot}). The red band corresponds to
the $\pm 1\sigma$ statistical uncertainty. The simulation method is
MC--II. In Figure 9a no cut is applied on the weight. In Figure 9b
a cut is applied at $\left|q_{w}\right|>0.1$.}
\label{Flo:ev_real_misID}
\end{figure}

The aim of the second investigation is to estimate the impurity, $\eta_{{\rm SM}}$,
for muons crossing one complete spectrometer ($n_{{\rm CMU}}=2$). For that
purpose, the data sample with particles crossing both spectrometers
is used ($n_{{\rm CMU}}=4$). It has been verified that all four CMU have
equal systematics \cite{Agafonova:2014zia, agafonova2010} and therefore
an equal impurity $\eta_{{\rm SM}}$ is expected for both spectrometers.
The fraction of tracks with a different charge sign measurement in
both spectrometers is then given by

\begin{equation}
\frac{n^{+-}}{n}=2\eta_{{\rm SM}}\left(1-\eta_{{\rm SM}}\right),\label{eq:mis_ID}
\end{equation}
\textcolor{black}{where $n$ is the total number of muons and $n^{+-}$
is the number of events where different charge signs are obtained.
The charge sign is reconstructed independently
in each spectrometer.
Figure~\ref{Flo:ev_real_misID}
shows the impurity obtained by this method  in the muon charge determination, as a function of momentum,
for real data and for MC predictions. The impurity is kept below 0.5\% and the muon
charge is correctly determined with an efficiency larger than 99.5\%
for momenta below 15 GeV/c, the momentum range relevant for the study
of $\nu_{\mu}\rightarrow\nu_{\tau}$ oscillations (Figure \ref{Flo:NoWeightCut}).
An additional cut at $\left|q_{w}\right|>0.1$ allows a further
reduction of the impurity at the cost of an increase of the fraction
of sign indetermination (Figure \ref{Flo:WeightCut}). This result
is relevant for most of the OPERA events as in 80\% of the CC interactions,
the muon crosses at least one full spectrometer. In 35\% of the cases
both spectrometers are crossed allowing a further reduction of
the impurity.}

\section{Conclusion and outlook\label{sec:Conclusion}}

\textcolor{black}{A new method (AMM) has been developed in the framework
of the OPERA experiment to improve the determination of the muon charge
sign in the spectrometers. In each CMU a weight is assigned
to the matching between the measured angles of the two straight track
segments at the entry and exit of the magnet arm, projected in the
plane of curvature. 
These weights are then combined to compute a charge sign estimator, the
sign of which determines the charge sign of the muon. Its modulus
measures the quality of this determination. A lower cut applied to
the weight improves the purity -- the fraction of correct charge sign
determinations -- at the cost of some reduction in the efficiency,
the fraction of muons for which the charge is determined. The purity
naturally increases with the number of CMUs that are crossed by
the muon. It has only a small momentum dependence and is affected
by the two irreducible effects: a) at small momentum, the multiple Coulomb
scattering suffered by the muon inside the magnet, and b) at high momentum
and small deflection,  the finite resolution in the measurement
of the track segment angles. }

\textcolor{black}{ AMM has been used to evaluate the purity in
the charge sign measurement by one spectrometer for real CNGS beam
data by comparing how often both spectrometers measure the same sign.
}The impurity is kept below 0.5\% in the momentum range relevant for
the OPERA main analysis. 

AMM analysis is part of a campaign of studies to reduce the
backgrounds for the OPERA experiment.
The estimation of the large--angle muon scattering background which was
formerly based on conservative assumptions has been recently updated\cite{ref:longhin}
and reduced to a level where charm decays now contribute for
about 95\% of the background in the muonic channel. This allows for a
more significant role of the AMM algorithm in improving the
sensitivity of the experiment.

The use of AMM is not, a priori, restricted to the configuration of the OPERA
spectrometers and could be  adapted to other experiments, for
example to decrease the fraction of CC events with wrong muon charge
sign determination or to better control the systematics in the separation
between\textcolor{black}{{} $\nu_{\mu}$ and $\bar{\nu}_{\mu}$} CC
interactions. Experiments with point--like track measurements in the
magnetic field (e.g.\textcolor{black}{{} \cite{michael2008}}) may adapt
the method for their purposes by using three points to form the triangle
shown in Figure~\textcolor{black}{\ref{Flo:triangle}} instead of
two tangents. 

\appendix

\section*{Appendix A}

\renewcommand{\theequation}{A.\arabic{equation}}The OSM used so far by the OPERA experiment for the charge sign determination
is described in detail in \citep{zimmermann2009}.
In each CMU$_i$ a weight $w_{i}$ is computed that takes account
the measurement precision of the angles $\phi_{1}$ and $\phi_{2}$
made by the two track segments with the transverse direction X
in the horizontal plane of projection, the plane of curvature,

\begin{equation}
w_{i}=\frac{\phi_{2}-\phi_{1}}{\sqrt{\sigma_{\phi_{1}}^{2}+\sigma_{\phi_{2}}^{2}}}
\end{equation}
 If all signs are equal, that sign represents the result. The weights
of each measurement are added in quadrature to form a global weight
for the final result. If the signs differ, the sign measured by the majority
of the CMU is used. Only the weights of the stations belonging to
the majority are used and added up for the output. If an equal number
of positive and negative signs is obtained, it is assumed, that the
measurement is disturbed by the presence of additional hits due to
the leakage of the tail of the hadronic and electromagnetic showers
from the target. Since showers are rapidly absorbed by the magnet, the
sign from the CMU closest to the main event vertex is rejected. 

\appendix

\section*{Appendix B}

\renewcommand{\theequation}{B.\arabic{equation}}

The quantities defined at the beginning of Sec.~\ref{subsec:id}
satisfy the Hessian normal form (see also Figures \ref{Flo:ChargeSignDet}
and \ref{Flo:triangle}) 

\begin{equation}
\left(\left(\begin{array}{c}
x_{j}\\
z_{j}
\end{array}\right)-d_{0,j}\left(\begin{array}{c}
\sin\phi_{j}\\
-\cos\phi_{j}
\end{array}\right)\right)\cdot\left(\begin{array}{c}
\sin\phi_{j}\\
-\cos\phi_{j}
\end{array}\right)=0\label{eq:Hesse-1-1},
\end{equation}
$\Delta\phi=\left|\phi_{2}-\phi_{1}\right|$ being the deflection angle.
If energy loss and multiple scattering are neglected the charged particle
trajectory in the magnetic field is an arc of a circle tangent to
both track segments at their magnet entry/exit points. 

One defines the quantities projected in the bending horizontal plane: 
\begin{itemize}
\item $\mathbf{L}$, a direction vector of the line connecting $A$, the
entry point of the upstream segment in the magnet and $B$, the exit
point of the downstream segment. 
\item $\mathbf{a}$, $\mathbf{b}$, a direction vector along the upstream segment and 
a direction vector opposite to the downstream segment, respectively,
\end{itemize}
If follows that

\begin{eqnarray}
\mathbf{b} & = & \left(\begin{array}{c}
z_{2}-z_{1}\\
x_{1}\left(z_{2}\right)-x_{1}\left(z_{1}\right)
\end{array}\right)\\
\mathbf{a} & = & \left(\begin{array}{c}
z_{1}-z_{2}\\
x_{2}\left(z_{1}\right)-x_{2}\left(z_{2}\right)
\end{array}\right)\\
\mathbf{L} & = & \left(\begin{array}{c}
z_{2}-z_{1}\\
x_{2}\left(z_{2}\right)-x_{1}\left(z_{1}\right)
\end{array}\right),
\end{eqnarray}
where $x_{j}\left(z_{i}\right)$, defined by Eq. (\ref{eq:Hesse-1-1}),
is the X coordinate of the intersection point of track segment
$j=1,2$ with respectively the front face of the magnet arm at $z=z_{1}$
and the back face at $z=z_{2}$. It also follows

\begin{eqnarray}
\alpha_{1} & = & \arccos\left(\frac{\mathbf{b}\cdot\mathbf{L}}{b\cdot L}\right)\cdot s_{1}\label{a-1}\\
\alpha_{2} & = & \arccos\left(-\frac{\mathbf{a}\cdot\mathbf{L}}{a\cdot L}\right)\cdot s_{2}\label{eq:-1},
\end{eqnarray}
where signs $s_{1,2}$ are given by

\begin{equation}
s_{1,2}=\frac{x_{1,2}\left(z_{2,1}\right)-x_{2,1}\left(z_{2,1}\right)}{|x_{1,2}\left(z_{2,1}\right)-x_{2,1}\left(z_{2,1}\right)|},
\end{equation}
In order to calculate both angles ($\alpha_{1}$ and $\alpha_{2}$),
one uses the second parameter of the track segment fit ($d_{0}$), usually neglected
for the charge sign determination. Assuming a perfect circular
trajectory in the magnet (no energy loss, no multiple Coulomb scattering,
homogeneous field) and infinite measurement precision, one get $\alpha_{1}=\alpha_{2}$
and $\Delta\phi=\alpha_{1}+\alpha_{2}$.

\section*{Acknowledgements}

We thank CERN for the successful operation of the CNGS facility and
INFN for the continuous support given to the experiment through its
LNGS laboratory. We acknowledge funding from our national agencies:
Fonds de la Recherche Scientifique--FNRS and Institut Interuniversitaire
des Sciences Nucl\'eaires for Belgium, MoSES for Croatia, CNRS and
IN2P3 for France, BMBF for Germany, INFN for Italy, JSPS, MEXT, QFPU
\linebreak (Global COE programme of Nagoya University) and Promotion
and Mutual Aid Corporation for Private Schools of Japan for Japan,
SNF, the University of Bern and ETH Zurich for Switzerland, the Russian
Foundation for Basic Research (grant 12--02--12142 ofim, 15--02--01056\_a),
the Program for the Support of Leading Scientific School, contract
SS 3110.2014.2, the Presidium of the Russian Academy of Sciences basic
research program Fundamental Properties of Matter and Astrophysics,
and the Ministry of Education and Science of the Russian Federation
for Russia, and the National Research Foundation of Korea Grant No.
2011--0029457 for Korea. The authors would also like to thank Laura
Vanhoefer for fruitful discussions and comments. B.B. and M.M. would
also like to thank Ann Fielding for language support.

\bibliographystyle{elsarticle-num}
\bibliography{ChargeWeight_References}

\end{document}